\documentclass[pra,aps,twocolumn,tightlines]{revtex4}

\usepackage[]{times,amsmath,epsfig,graphicx}

\begin{document}

\title{Stability of Resonant Opto-Mechanical Oscillators}

\author{A. B. Matsko, A. A. Savchenkov, and L. Maleki}

\affiliation{OEwaves Inc., 465 N. Halstead Street, Ste. 140, Pasadena, CA 91107}

\begin{abstract}
We theoretically study the frequency stability of an opto-mechanical radio frequency oscillator based on resonant interaction of two optical and one mechanical modes of the same optical microcavity. A generalized expression for the phase noise of the oscillator is derived using Langevin formalism and compared to the phase noise of existing electronic oscillators.
\end{abstract}


\maketitle

\section{Introduction}

The opto-mechanical oscillator (OMO) generates spectrally pure radio frequency (RF) signals \cite{hosseinzadeh06pra,rokhsari06apl,kippenberg07oe,kippenberg08s,hosseinzadeh08apl,hosseinzadeh10jstqe} due to ponderomotive interaction between photons and phonons. Phase noise and linewidth are the main characteristics  detremining the performance of an oscillator  and its practical usefulness. It was shown that the OMO linewidth can be small enough to be ultimately described by a Schawlow-Townes-like formula \cite{vahala08pra}. A Leeson model \cite{leeson66pieee} of the phase noise far from the carrierfor a radiation pressure driven OMO  was presented in \cite{tallur10ifcs}. In this paper we derive a generalized formula for the OMO phase noise that takes into account the noise of the light pumping the resonator. Using the formula, we analyze Allan deviation of the oscillator frequency as well as its linewidth. The predicted performance of the OMO is then compared with the performance of existing electronic oscillators.

Opto-mechanical oscillation can be described as a strongly nondegenerate parametric process in which a pump photon is transformed to a photon of lower frequency (Stokes photon) and a phonon \cite{matsko02pra}.  This occurs when the pump power exceeds a certain threshold determined by the loss of the system, and the coupling efficiency between the light and the mechanical modes. In the case of microcavity-based OMO, the process is allowed if the resonator has at least two optical modes with frequency equal to the frequency of a mechanical mode. This condition satisfies energy conservation. The momentum conservation (phase matching of the parametric interaction) is met by the requirement of spatial overlap between the optical and the mechanical modes that ensure the convolution integral of the electric fields and the mechanical displacement, as well as mechanical strain, is nonzero \cite{matsko09prl}.

Generally, the radiation by a laser is assumed to be phase insensitive. It means that the phase of the emitted light in a laser does not depend on the phase of the pump, and that there is no phase matching conditions (preferred directionality) in the system. An OMO, even one based on Brillouin lasing \cite{grudinin09prl,tomes09prl},  has different properties. The mechanical quality factor is high enough to endow the generated phonons with some well defined phase. The phase of the Stokes light emitted by the OMO depends on the phonon phase, so the amplification of the Stokes wave becomes phase sensitive and the phase fluctuations of the optical pump leak to the phase fluctuations of the Stokes light. In what follows we develop a model for an OMO and find the phase noise of the generated signal.

It was shown that the linewidth of an OMO increases if generation of an anti-Stokes optical sideband is allowed by the system architecture \cite{matsko02pra,vahala08pra}. In this paper we analyze an ideal case where only two high quality (Q-) factor optical modes and a single mechanical mode interact. The optical pump is resonant with higher frequency optical mode and the Stokes optical sideband is generated in the lower frequency mode. The bandwidth of the optical modes is assumed to be much smaller than the mechanical frequency. This kind of interaction has been realized in oscillators based on stimulated Brillouin scattering \cite{grudinin09prl,tomes09prl} and surface acoustic waves \cite{bahl11nc,savchenkov11ol}.

We write three coupled Langevin equations for the modes and solve them analytically to evaluate the phase noise of the OMO. To obtain a consistent result, we take into account the phase noise of the laser used for pumping the OMO. We show that one limitation of an OMO is that the frequency stability of the mechanical mode is determined  in the same way as the electronic quartz oscillators. The fluctuations of the thermal bath as well as the thermodynamic noise and drift of the mechanical mode limit the frequency stability. The pumping light is used as the power source in an OMO, similar to the electric power source in electronic oscillators and, therefore, it is unlikely that the OMO will outperform its electronic analogs, unless the frequency bandwidth of the optical modes is significantly smaller than the frequency bandwidth of the mechanical mode. An advantage of an OMO, on the other hand, is in its potential long term stabilization via stability transfer from the optical frequency domain.

The paper is organized as follows. The equations are presented in Section II, and their solutions are described in Section III. The OMO phase noise and linewidth are analyzed and compared to similar parameters of existing quartz oscillators in Section IV.

\section{Basic Equations}

The triply-resonant opto-mechanical interaction is described by equations
\begin{eqnarray}
\label{q1} && \dot  A = -\Gamma_A A - ig C B +  F_A,
\\
\label{q2} && \dot {B} = - \Gamma_B   B -i g C^\dag A + F_B,
\\
\label{q3} && \dot{C} = -  \Gamma_C   C -  i g B^\dag A + F_C.
\end{eqnarray}
where $  A$, $  B$, and $  C$ are the slowly-varying
amplitudes the pump (optical), the Stokes (optical mode red shifted with respect to the pump), and the signal (mechanical) fields; $\Gamma_A$, $\Gamma_B$, and  $\Gamma_C$ are the linear resonant terms of optical and mechanical modes respectively
\begin{eqnarray} \nonumber
&& \Gamma_{A} =  i (\omega_a -\omega_0) + \gamma + \gamma_{ca},
\\
\nonumber && \Gamma_{B} =  i (\omega_b - \omega_-) + \gamma + \gamma_{cb},
\\
\nonumber && \Gamma_C = i (\omega_c - \omega_M) + \gamma_M,
\end{eqnarray}
$\gamma$ and $\gamma_M$ are the intrinsic decay rates of the optical and mechanical modes,
$\gamma_{ca}$ and $\gamma_{cb}$ are optical loading (coupling) rates (the loading of the optical modes can be different because modes belong to different families); $g$ is the opto-mechanical coupling constant,
\begin{equation}
g=\omega_0\sqrt{\frac{K_\epsilon \hbar }{2m^*L^2 \omega_c}},
\end{equation}
$K_\epsilon$ is the correction coefficient showing that radiation pressure results not only in a change in the size of the resonator, but also in its index of refraction through strain, $m^*$ is an effective mass of the mechanical mode, $L$ is an effective spatial parameter of the mode.

The terms $ F_A$ and $ F_{B}$ represent  Langevin forces with two uncorrelated parts arising from the internal and coupling loss of the modes
\begin{eqnarray}
 F_A =  F_{cA}+ F_{rA}, & & \langle  F_A \rangle = e^{i\phi_{F_A}}\sqrt{\frac{2P\gamma_{ca}}{\hbar \omega_0 }}, \\
\langle  F_{cA}(t) F_{cA}^\dag (t') \rangle &=&
2\gamma_{ca} \delta (t-t'), \\
\langle  F_{rA}(t) F_{rA}^\dag (t') \rangle &=&
2\gamma \delta (t-t'),
\end{eqnarray}
\begin{eqnarray}
 F_B =  F_{cB}+ F_{rB}, & & \langle  F_B \rangle = 0, \\
\langle  F_{cB}(t) F_{cB}^\dag (t') \rangle &=&
2\gamma_{cb} \delta (t-t'), \\
\langle  F_{rB}(t) F_{rB}^\dag (t') \rangle &=&
2\gamma \delta (t-t'),
\end{eqnarray}
where $\langle \dots \rangle$ stands for ensemble averaging,
and $P$ is the power of the external optical pump of the mode $A$.

The Langevin force describing the thermal
fluctuations of the mechanical system is defined in the similar way
\begin{eqnarray}
\langle  F_{C} \rangle = 0, \\
\langle  F_{C}(t) F_{C}^\dag (t') \rangle &=&
2\gamma_{M} (\bar
{n}_{th}+1) \delta (t-t')
\end{eqnarray}
where $\bar {n}_{th} = [\exp(\hbar \omega_c/k_BT) -1]^{-1}$ is the averaged number of
thermal phonons leaking from the thermal bath of temperature $T$ to the mechanical mode, $k_B$ is the Boltzmann constant.

We present the slowly-varying complex amplitudes as
\begin{eqnarray}
  A  = | A | e^{i \phi_A}, \\
  B  = | B | e^{i \phi_B}, \\
  C  = | C | e^{i \phi_C},
\end{eqnarray}
and derive two sets of equations for the amplitude and phase parts of the complex amplitudes from Eqs.~(\ref{q2}-\ref{q3})
\begin{eqnarray} \label{am1}
&& | \dot A | = -(\gamma + \gamma_{ca}) |A|- g |C||B|\sin \phi+ \\ \nonumber && |\langle  F_A \rangle| \cos (\phi_{F_A}-\phi_A)+F_{Ar}, \\
&& \label{am2} | \dot B | =-(\gamma + \gamma_{cb}) |B|+ g |C||A|\sin \phi+F_{Br}, \\
&& \label{am3} | \dot C | = -\gamma_M |C| + g |B||A|\sin \phi+F_{Cr};
\end{eqnarray}
and
\begin{eqnarray}
&& \label{ph1} \dot \phi_A  = -(\omega_a -\omega_0)- g \frac{|C||B|}{|A|}\cos \phi+ \\ \nonumber &&\frac{|\langle  F_A \rangle|}{|A|} \sin (\phi_{F_A}-\phi_A)+ \frac{F_{Ai}}{|A|}, \\
&& \label{ph2} \dot \phi_B  = -(\omega_b - \omega_-)-g \frac{|C||A|}{|B|}\cos \phi+\frac{F_{Bi}}{|B|}, \\
&& \label{ph3} \dot \phi_C  = -(\omega_c - \omega_M)-g \frac{|B||A|}{|C|}\cos \phi+\frac{F_{Ci}}{|C|};
\end{eqnarray}
where
\begin{eqnarray}
&& \phi=\phi_A-\phi_B-\phi_C, \\
&& F_{Ar}= \frac{1}{2} (F_{cA}e^{-i \phi_A} +F_{cA}^\dag e^{i \phi_A})+\\ \nonumber && \frac{1}{2} (F_{rA}e^{-i \phi_A}+F_{rA}^\dag e^{i \phi_A}), \\
&& F_{Br}=\frac{1}{2} (F_{cB}e^{-i \phi_B}+F_{cB}^\dag e^{i \phi_B})+\\ \nonumber && \frac{1}{2} (F_{rB}e^{-i \phi_B}+F_{rB}^\dag e^{i \phi_B}), \\
&& F_{Cr}=\frac{1}{2} (F_{C}e^{-i \phi_C}+F_{C}^\dag e^{i \phi_C}), \\
&& F_{Ai}= \frac{1}{2i} (F_{cA}e^{-i \phi_A}-F_{cA}^\dag e^{i \phi_A})+\\ \nonumber && \frac{1}{2i} (F_{rA}e^{-i \phi_A}-F_{rA}^\dag e^{i \phi_A}), \\
&& F_{Bi}=\frac{1}{2i} (F_{cB}e^{-i \phi_B}-F_{cB}^\dag e^{i \phi_B})+\\ \nonumber && \frac{1}{2i} (F_{rB}e^{-i \phi_B}-F_{rB}^\dag e^{i \phi_B}), \\
&& F_{Ci}=\frac{1}{2i} (F_{C}e^{-i \phi_C}-F_{C}^\dag e^{i \phi_C}).
\end{eqnarray}

Equations (\ref{am1}-\ref{ph3}) completely describe the triply-resonant opto-mechanical interaction.

\section{Solution}

Using the set of equations for the phase (\ref{ph1}-\ref{ph3}) we get the equation for phase difference $\phi$
\begin{eqnarray}
\nonumber && \dot \phi = -(\omega_a-\omega_b-\omega_c)+g \cos\phi \left ( \frac{|B||A|}{|C|}+ \frac{|C||A|}{|B|} -  \frac{|C||B|}{|A|}\right )+ \\ && \frac{|\langle  F_A \rangle|}{|A|} \sin (\phi_{F_A}-\phi_A)+ \frac{F_{Ai}}{|A|} - \frac{F_{Bi}}{|B|}-\frac{F_{Ci}}{|C|}, \label{phi1}
\end{eqnarray}
We introduce $\psi = \phi-\pi/2$ and rewrite Eq.~(\ref{phi1}) as
\begin{eqnarray}
\nonumber && \dot \psi = -(\omega_a-\omega_b-\omega_c)-g \sin \psi \left ( \frac{|B||A|}{|C|}+ \frac{|C||A|}{|B|} -  \frac{|C||B|}{|A|}\right ) \\ && +\frac{|\langle  F_A \rangle|}{|A|} \sin (\phi_{F_A}-\phi_A)+ \frac{F_{Ai}}{|A|} - \frac{F_{Bi}}{|B|}-\frac{F_{Ci}}{|C|}.
\end{eqnarray}
Next, we introduce an expectation time-independent value for the phase difference, $\langle \psi \rangle$,  and fluctuational time-dependent part, $\delta \phi$, so that $\psi=\langle \psi \rangle + \delta \phi$, and get
\begin{eqnarray}
\nonumber  && (\omega_a-\omega_b-\omega_c)+\\ \nonumber && g \sin \langle \psi \rangle \left ( \frac{|B||A|}{|C|}+ \frac{|C||A|}{|B|} -  \frac{|C||B|}{|A|}\right )=\\ && \frac{|\langle  F_A \rangle|}{|A|} \sin (\phi_{F_A}-\phi_A), \label{psi}\\ \nonumber &&
\delta \dot \phi + g \delta \phi \left ( \frac{|B||A|}{|C|}+ \frac{|C||A|}{|B|} -  \frac{|C||B|}{|A|}\right )=\\ \nonumber && \delta \left [ -g \sin \langle \psi \rangle \left ( \frac{|B||A|}{|C|}+ \frac{|C||A|}{|B|} -  \frac{|C||B|}{|A|}\right ) + \right. \\  && \left. \frac{|\langle  F_A \rangle|}{|A|} \sin (\phi_{F_A}-\phi_A) \right ]  +\frac{F_{Ai}}{|A|} - \frac{F_{Bi}}{|B|}-\frac{F_{Ci}}{|C|}, \label{phi}
\end{eqnarray}
where $\delta[\dots ]$ means a deviation from the expectation value.

Using Eqs.~(\ref{am1}-\ref{ph3}) we find several relationships for the expectation values of the oscillator parameters in the steady state
\begin{eqnarray}&&
\frac{|   B  |^2}{|   C  |^2} = \frac{\gamma_M}{\gamma + \gamma_{cb}}, \label{s1} \\ &&
|   A  |^2= \frac{\gamma_M}{\gamma + \gamma_{cb}} \frac{|\Gamma_B|^2}{g^2} \label{s2} \\ && \frac{\omega_b - \omega_- }{\omega_c - \omega_M } = \frac{\gamma + \gamma_{cb}}{\gamma_M}, \label{omm} \\
&& e^{i\langle \psi \rangle }=e^{i \phi_{\Gamma_B}}. \label{psiav}
\end{eqnarray}

The expectation value of the amplitude of the field in the pumped mode increases below the oscillation threshold with increase of the pump power, in accordance with  $   |A| =   |F_A| /|\Gamma_A|$, and then stays constant, in accordance with Eq.~(\ref{s2}).

For the sake of simplicity, we assume that the Stokes sideband has much lower power compared with the pump, $|A| \gg |B|$, and that the system is triply-resonant (the expectation values of the frequencies are resonant with corresponding modes). We note that
\begin{eqnarray}
\frac{|B||A|}{|C|}+ \frac{|C||A|}{|B|} -  \frac{|C||B|}{|A|} = -  \frac{g}{|\Gamma_B|} |B|^2+ \\ \nonumber \left (  \sqrt{\frac{\gamma_M}{\gamma + \gamma_{cb}}} +\sqrt{\frac{\gamma + \gamma_{cb}}{\gamma_M}} \right ) \sqrt{\frac{\gamma_M}{\gamma + \gamma_{cb}} } \frac{|\Gamma_B|}{g} ,
\end{eqnarray}
so that for the  case of relatively weak Stokes sideband and all-resonant tuning
\begin{eqnarray} \label{simplificat}
g  \left ( \frac{|B||A|}{|C|}+ \frac{|C||A|}{|B|} -  \frac{|C||B|}{|A|}\right ) \simeq
\gamma_M + \gamma + \gamma_{cb}.
\end{eqnarray}
Using the assumptions and (\ref{simplificat}) we find
\begin{eqnarray}
\delta \left [ -g \sin \langle \psi \rangle \left ( \frac{|B||A|}{|C|}+ \frac{|C||A|}{|B|} -  \frac{|C||B|}{|A|}\right ) + \right. \\ \nonumber \left. \frac{|\langle  F_A \rangle|}{|A|} \sin (\phi_{F_A}-\phi_A) \right ] \approx
(\delta \phi_{F_A}-\delta \phi_A) \frac{|\langle  F_A \rangle|}{|A|},
\end{eqnarray}
where we took into account that, in accordance with Eq.~(\ref{psi}) and Eq.~(\ref{psiav}), $\langle \phi_{F_A}-\phi_A \rangle = 0$ for the resonant tuning.

We obtain a set of linear equations consisting of Eqs.~(\ref{ph1}) and (\ref{phi})
\begin{eqnarray} \label{dotphi}
&& \delta \dot \phi + (\gamma_M + \gamma + \gamma_{cb}) \delta \phi = \\ \nonumber && (\delta \phi_{F_A}-\delta \phi_A) \frac{|\langle  F_A \rangle|}{|A|} +\frac{F_{Ai}}{|A|} - \frac{F_{Bi}}{|B|}-\frac{F_{Ci}}{|C|}, \\
&& \delta \dot \phi_A  = (\delta \phi_{F_A}-\delta \phi_A) \frac{|\langle  F_A \rangle|}{|A|}+ \frac{F_{Ai}}{|A|}. \label{dotphiA}
\end{eqnarray}
The condition of small Stokes sideband can be written in the form
\begin{equation} \label{force1}
\frac{|\langle  F_A \rangle|}{|A|} = \gamma + \gamma_{ca}.
\end{equation}

Substituting Eq.~(\ref{force1}) into (\ref{dotphi}) and (\ref{dotphiA}) we derive
\begin{eqnarray} \label{delphia1}
&& \delta \dot \phi_A + (\gamma + \gamma_{ca}) \delta  \phi_A = \frac{F_{Ai}}{|A|}, \\ \label{delphia2}
&& \delta \dot \phi_B + \delta \dot \phi_C + (\gamma_M + \gamma + \gamma_{cb})(\delta \phi_B + \delta \phi_C ) = \\ \nonumber && (\gamma_M + \gamma + \gamma_{cb})\delta \phi_A+\frac{F_{Bi}}{|B|}+\frac{F_{Ci}}{|C|}.
\end{eqnarray}
The third equation for  phase deviations can be derived from Eqs.~(\ref{ph2}) and (\ref{ph3}):
\begin{eqnarray} \nonumber
\delta \dot \phi_B \sqrt{\frac{\gamma_M}{\gamma + \gamma_{cb}}}-\delta \dot \phi_C \sqrt{\frac{\gamma + \gamma_{cb}}{\gamma_M}}= \\ \sqrt{\frac{\gamma_M}{\gamma + \gamma_{cb}}} \frac{F_{Bi}}{|B|}-\sqrt{\frac{\gamma + \gamma_{cb}}{\gamma_M}}\frac{F_{Ci}}{|C|}. \label{delphia3}
\end{eqnarray}

We solve the set of linear differential equations (\ref{delphia1}-\ref{delphia3}) using Fourier transform, e.g.
\begin{eqnarray}
F_{Bi}  &=& \int^{\infty}_{-\infty} f_{Bi}(\omega) e^{-i\omega t} \frac{d\omega}{2\pi}, \\
F_{Ci}  &=&  \int^{\infty}_{-\infty} f_{Ci}(\omega) e^{-i\omega t} \frac{d\omega}{2\pi},
\end{eqnarray}
where  $\hat f_{Bi}(\omega)$ and $\hat f_{Ci}(\omega)$ are the Fourier components of the noise,
\begin{eqnarray}
\langle  F_{Bi}(t) F_{Bi} (t') \rangle &=& \frac{1}{2}(\gamma+\gamma_{cb}) \delta (t-t'), \\
\langle  F_{Ci}(t) F_{Ci} (t') \rangle &=& \gamma_{M} \left (\bar {n}_{th}+\frac 1 2  \right ) \delta (t-t'), \\
\langle f_{Bi}(\omega)f_{Bi}(\omega') \rangle &=& \pi (\gamma+\gamma_{cb}) \delta (\omega+\omega'), \\
\langle f_{Ci}(\omega)f_{Ci}(\omega') \rangle &=& 2 \pi \gamma_{M} \left (\bar {n}_{th}+\frac 1 2  \right ) \delta (\omega+\omega').
\end{eqnarray}

Readout of the OMO signal is accomplished by tracking thradio frequency beat note produced by the optical pump and the generated sideband on a fast photodiode, so the phase and frequency of the measured oscillation signal are determined by the argument of the product of optical amplitudes ($AB^*$). The phase noise of the signal is given by difference $\delta \phi_B-\delta \phi_A$. On the other hand, the mechanical frequency can be read using electronics means, e.g. a capacitive displacement sensor. Neglecting by the electronics back action, we can estimate the phase noise of the signal evaluating $\delta \phi_C$. We find
expressions for Fourier amplitudes of these parameters
\begin{eqnarray}
\label{deltadelt0} && \delta  \phi_A (\omega)= \frac{f_{Ai}}{|A|} \frac{1}{-i \omega +\gamma+\gamma_{ca}}, \\
&& \label{deltadelta} \delta \phi_A(\omega) - \delta  \phi_B(\omega) = \\ \nonumber &&
\frac{-i \omega +\gamma_M}{-i \omega +\gamma_M+\gamma+\gamma_{cb}} \delta \phi_A (\omega) - \\ \nonumber &&
\frac{-i\omega+\gamma_M}{-i\omega(-i \omega +\gamma_M+\gamma+\gamma_{cb})} \frac{f_{Bi}}{|B|} +\\ \nonumber && \frac{\gamma+\gamma_{cb}}{-i\omega(-i \omega +\gamma_M+\gamma+\gamma_{cb})}\frac{f_{Ci}}{|C|}, \\
&& \label{deltaC} \delta \phi_C(\omega) = \frac{\gamma_M}{-i \omega +\gamma_M+\gamma+\gamma_{cb}} \delta \phi_A(\omega)-\\ \nonumber &&
\frac{\gamma_M}{-i\omega(-i \omega +\gamma_M+\gamma+\gamma_{cb})} \frac{f_{Bi}}{|B|} + \\ \nonumber && \frac{-i \omega +\gamma+\gamma_{cb}}{-i\omega(-i \omega +\gamma_M+\gamma+\gamma_{cb})}\frac{f_{Ci}}{|C|}. \nonumber
\end{eqnarray}

Equations (\ref{deltadelta}) and (\ref{deltaC}) can be used to find the phase noise of the OMO. For example, single-sideband phase noise ${\cal L}_c (\omega)$ of the mechanical oscillation is defined as,
\begin{equation}
\langle \delta \phi_C(t) \delta \phi_C(t-\tau) \rangle = \int \limits_{-\infty}^\infty {\cal L}_c (\omega) e^{i\omega t} \frac{d \omega}{2\pi}.
\end{equation}
Using definition
\begin{eqnarray}
\delta \phi_C(t) &=& \int^{\infty}_{-\infty} \delta \phi_C(\omega) e^{-i\omega t} \frac{d\omega}{2\pi},
\end{eqnarray}
we find
\begin{eqnarray} \label{lc}
&& {\cal L}_{c}= \\ \nonumber && \frac{\gamma_M^2}{\omega^2 +(\gamma_M+\gamma+\gamma_{cb})^2} \frac{\gamma_{ca}^2}{\omega^2 +(\gamma+\gamma_{ca})^2}{\cal L}_{in}+ \\ \nonumber && \frac{\gamma_M^2}{\omega^2[\omega^2 +(\gamma_M+\gamma+\gamma_{cb})^2]} \frac{\gamma+\gamma_{cb}}{2|B|^2}+ \\ \nonumber && \frac{\omega^2 +(\gamma+\gamma_{cb})^2}{\omega^2[\omega^2 +(\gamma_M+\gamma+\gamma_{cb})^2]} \frac{\gamma_{M}}{|C|^2}\left ( \bar {n}_{th}+\frac 1 2  \right ),
\end{eqnarray}
where ${\cal L}_{in}$ stands for Fourier frequency dependent input phase noise of the pump laser. This value is usually much larger than the contribution from the quantum white noise of the laser.

Using similar reasoning we obtain
\begin{eqnarray}\label{lab}
&& {\cal L}_{a-b}= \\ \nonumber && \frac{\omega^2 +\gamma_M^2}{\omega^2 +(\gamma_M+\gamma+\gamma_{cb})^2} \frac{\gamma_{ca}^2}{\omega^2 +(\gamma+\gamma_{ca})^2}{\cal L}_{in}+ \\ \nonumber && \frac{\omega^2+\gamma_M^2}{\omega^2[\omega^2 +(\gamma_M+\gamma+\gamma_{cb})^2]} \frac{\gamma+\gamma_{cb}}{2|B|^2}+ \\ \nonumber && \frac{(\gamma+\gamma_{cb})^2}{\omega^2[\omega^2 +(\gamma_M+\gamma+\gamma_{cb})^2]} \frac{\gamma_{M}}{|C|^2}\left ( \bar {n}_{th}+\frac 1 2  \right ).
\end{eqnarray}

It is also useful to write an expression for the phase noise of the optical Stokes mode, determining the stability and spectral purity of the Brillouin laser \cite{grudinin09prl}
\begin{eqnarray}\label{lab1}
&& {\cal L}_{b}= \\ \nonumber && \frac{(\gamma+\gamma_{cb})^2}{\omega^2 +(\gamma_M+\gamma+\gamma_{cb})^2} \frac{\gamma_{ca}^2}{\omega^2 +(\gamma+\gamma_{ca})^2}{\cal L}_{in} + \\ \nonumber && \frac{\omega^2+\gamma_M^2}{\omega^2[\omega^2 +(\gamma_M+\gamma+\gamma_{cb})^2]} \frac{\gamma+\gamma_{cb}}{2|B|^2}+ \\ \nonumber && \frac{(\gamma+\gamma_{cb})^2}{\omega^2[\omega^2 +(\gamma_M+\gamma+\gamma_{cb})^2]} \frac{\gamma_{M}}{|C|^2}\left ( \bar {n}_{th}+\frac 1 2  \right ).
\end{eqnarray}

Formulas (\ref{lc}), (\ref{lab}) and (\ref{lab1}) can be simplified further using the ratio between the photon number in the Stokes mode and the phonon number in the mechanical mode (\ref{s1}), and the expression connecting the number of pump and Stokes photons and output power of the pump and Stokes light
\begin{eqnarray}
|A|^2=\frac{2\gamma_{ca}}{(\gamma+\gamma_{ca})^2} \frac{P}{\hbar \omega_0}, \;\;
|B|^2=\frac{P_{Bout}}{2 \gamma_{cb} \hbar \omega_0},
\end{eqnarray}
where we assumed that the carrier frequency of the Stokes light is approximately equal to the frequency of the pump light. Finally we get
\begin{eqnarray}
&& \label{pn1} {\cal L}_{a-b}= \\ \nonumber && \frac{\omega^2 +\gamma_M^2}{\omega^2 +(\gamma_M+\gamma+\gamma_{cb})^2} \frac{\gamma_{ca}^2}{\omega^2 +(\gamma+\gamma_{ca})^2}{\cal L}_{in} + \\ \nonumber &&
\frac{\omega^2+2\gamma_M^2(\bar {n}_{th}+1)}{ [\omega^2 +(\gamma_M+\gamma+\gamma_{cb})^2]} \frac{\gamma_{cb}(\gamma+\gamma_{cb})}{\omega^2} \frac{\hbar \omega_0}{P_{Bout}}, \\
&& \label{pn2} {\cal L}_{c}= \\ \nonumber && \frac{\gamma_M^2}{\omega^2 +(\gamma_M+\gamma+\gamma_{cb})^2} \frac{\gamma_{ca}^2}{\omega^2 +(\gamma+\gamma_{ca})^2}{\cal L}_{in}+ \\ \nonumber &&
\frac{\gamma_M^2\left \{ \bar {n}_{th}+1+ \omega^2/[2(\gamma+\gamma_{cb})^2] \right \}}{[\omega^2 +(\gamma_M+\gamma+\gamma_{cb})^2]} \frac{2 \gamma_{cb}(\gamma+\gamma_{cb})}{\omega^2} \frac{\hbar \omega_0}{P_{Bout}}, \\
&& \label{pn3} {\cal L}_{b}= \\ \nonumber && \frac{(\gamma+\gamma_{cb})^2}{\omega^2 +(\gamma_M+\gamma+\gamma_{cb})^2} \frac{\gamma_{ca}^2}{\omega^2 +(\gamma+\gamma_{ca})^2}{\cal L}_{in} + \\ \nonumber &&
\frac{\omega^2+2\gamma_M^2(\bar {n}_{th}+1)}{ [\omega^2 +(\gamma_M+\gamma+\gamma_{cb})^2]} \frac{\gamma_{cb}(\gamma+\gamma_{cb})}{\omega^2} \frac{\hbar \omega_0}{P_{Bout}}. \\
\end{eqnarray}

Equation (\ref{pn1}) gives a complete description of the phase noise characteristic of the radio frequency photonic oscillator based on demodulation of the light output of the OMO on a fast photodiode. Equation (\ref{pn2}) shows the limiting phase noise of the oscillating mechanical mode that could observed, if the mechanical oscillation is measured using an external devise that does not disturb the system. Equation (\ref{pn3}) describes the phase noise of the SBS laser. These expressions can be used to find the linewidth and Allan deviation of the corresponding OMO signals. By definition, the Allan variance of the frequency of the oscillator  is given by
\begin{equation} \label{sigm}
\sigma^2(\tau)=\int \limits_{0}^{\infty} \frac{4\omega^2 {\cal L}}{\omega_0^2} \frac{\sin^4(\omega \tau/2)}{(\omega \tau/2)^2} \frac{d\omega}{2\pi}.
\end{equation}

Let us consider the case where the OMO signal is retrieved by demodulation of the light leaving the resonator on a  photodiode. The single sided power density of the phase noise is related to the linewidth of the oscillator $\Delta \nu$ as $ {\cal L}(\omega \rightarrow 0)= 2 \pi \Delta\nu/\omega^2$. We find for the cases of relatively low and relatively high-Q of the mechanical mode
\begin{eqnarray}
\nonumber && \Delta \nu_{a-b} |_{\gamma_M\gg\gamma} \simeq  \frac{\gamma_{ca}^2}{(\gamma+\gamma_{ca})^2} \Delta \nu_{pump}+ \\  && \gamma_{cb}(\gamma+\gamma_{cb}) (\bar {n}_{th}+1) \frac{\hbar \omega_0}{\pi P_{Bout}}, \\
\nonumber  && \Delta \nu_{a-b} |_{\gamma_M\ll\gamma} \simeq  \frac{\gamma_M^2}{(\gamma+\gamma_{cb})^2} \frac{\gamma_{ca}^2}{(\gamma+\gamma_{ca})^2} \Delta \nu_{pump}+ \\  && \frac{\gamma_{cb}}{\gamma+\gamma_{cb}} \gamma_M^2 (\bar {n}_{th}+1) \frac{\hbar \omega_0}{\pi P_{Bout}}.
\end{eqnarray}
Therefore, in the case of low-Q mechanical mode, the linewidth of the radio frequency beat note generated by the OMO  is determined by the linewidth of the pumping light; while in the case of high-Q mechanical mode, the linewidth is described  by Schawlow-Townes-like formula \cite{vahala08pra} and the phase noise of the pumping light is suppressed.

\section{Discussion}

Let us estimate the phase noise of the oscillator and compare it to the phase noise of an electronic oscillator. We assume that $Q_M=10^5$, $\omega_M=2\pi \times 10^8$~rad$/$s, $\gamma_M=2\pi \times 500$~rad$/$s, $\gamma=2\pi \times 10^4$~rad$/$s, $\gamma_{cb}=\gamma_{ca}=2\pi \times 10^5$~rad$/$s, $\omega_0=2 \pi \times 10^{14}$~rad$/$s, $\Delta \nu_{pump}=1$~kHz, $P_{Bout}=100\ \mu$W, and $P=1$~mW.

The light escaping the cavity is demodulated on a photodiode to produce the radio frequency signal. The photodiode introduces thermal noise and white shot noise
\begin{equation} \label{thermal}
{\cal L}_{PD} = \frac{2qR \rho P_D + k_BT}{P_{RF}}
\end{equation}
in addition to the phase noise (\ref{pn1}) coming from the opto-mechanical oscillator. Here $\rho$ is the resistance of the photodiode, $R=\eta q/\hbar \omega_0$ is the responsivity of the photodiode, $\eta$ is quantum efficiency of the photodiode, $q$ is the electron charge, and $P_D$ is the total optical power reaching the photodiode. We assume that $P_D=P$. The thermal noise depends on the ambient temperature $T$ and the power of the radio frequency signal leaving the photodiode $P_{RF}=2 \rho R^2 P P_{Bout}$. To find its value we assume that the resistance at the output of the photodiode is $\rho=50$~Ohm, the responsivity of the photodiode is $0.8$~A$/$W, and the temperature is $T=300$~K. For these parameters, the expected power of the radio frequency signal escaping the photodiode is $P_{RF}=6\ \mu$W.

\begin{figure}[ht]
  \centering
  \includegraphics[width=8.5cm]{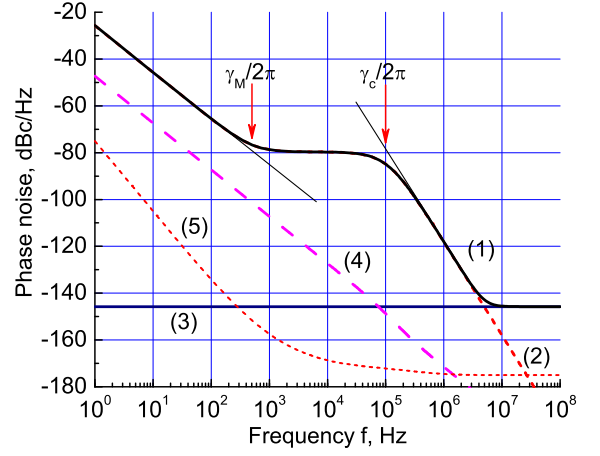}
\caption{ Phase noise of the opto-mechanical oscillator (1) characterized with parameters defined in the text. Curve (2) defines the contribution from the phase noise of the pumping laser. Curve (3) stands for the thermal and white shot noise defined in Eq.~(\ref{thermal}). Curve (4) shows phase diffusion of the opto-mechanical oscillator. Curve (5) stands for a commercially available 100~MHz ovenized quartz oscillator.} \label{fig1}
\end{figure}

Adding ${\cal L}_{PD}$ and ${\cal L}_{a-b}$ we plot the spectrum of the single sideband phase noise (line (1) in Fig.~(\ref{fig1})). It is dominated by the phase noise of the pump laser. If a very narrow linewidth laser is used instead, the phase noise of the OMO would be given by the convolution of curves (4) and (3) (phase diffusion of the oscillator as well as shot noise). In the same figure we also show the phase noise of a commercially available 100 MHz oven controlled quartz oscillator (line (5) in Fig.~(\ref{fig1})). Apparently, the electronic oscillator has much lower phase noise than the OMO, even though thermal drifts resulting in the flicker noise are not taken into account in our analysis.

Let us evaluate the sources of frequency drift of the opto-mechanical oscillator. We note that, according to (\ref{omm}) and energy conservation law, $\omega_0=\omega_-+\omega_M$,
\begin{eqnarray}
\omega_0-\omega_-=\omega_M= \\ \nonumber \frac{\gamma+\gamma_{cb}}{\gamma+\gamma_{cb}+\gamma_M} \omega_c+ \frac{\gamma_M}{\gamma+\gamma_{cb}+\gamma_M} (\omega_0-\omega_b).
\end{eqnarray}
The OMO frequency drifts if the frequencies of the optical, $\omega_b$, and mechanical, $\omega_c$, modes drift.

It is possible to lock the frequency of the pump light to the frequency of the corresponding optical mode, $\omega_0=\omega_a$. Then the OMO frequency becomes dependent on the parameters of the microcavity only: $\omega_a-\omega_b$ as well as $\omega_c$. This is similar to the case of a quartz oscillator, except the OMO frequency depends on both the eigenfrequency of the mechanical and optical modes. The dependence of the oscillation on the drift of the mechanical mode can be suppressed if the bandwidth of the optical mode is much smaller than the bandwidth of the mechanical mode.

A clear advantage of the OMO over an electronic oscillator is that it can be readily stabilized by locking the pump frequency to the frequency of the corresponding cavity mode, $\omega_0=\omega_a$, and then locking the temperature of the cavity to a thermally insensitive optical reference line, for example, an atomic transition. Such a stabilization of the temperature will stabilize the long term drift of the oscillation \cite{matsko10ifcs}.

\section{Conclusion}
We have theoretically studied the phase noise of a triply-resonant opto-mechanical oscillator based on a nonlinear optical microcavity. The oscillator generates Stokes optical photons and mechanical phonons out of photons of a coherent pumping light.  We have shown that the spectral purity of the opto-mechanical signal is primarily limited by the phase noise of the pump laser. The overall short term performance of the oscillator is expected to be worse than the performance of a conventional electronic oscillator. An important advantage of the opto-mechanical oscillator is the possibility to optically stabilize it. With a proper electronic locking scheme it is possible to transfer frequency stability from the optical domain to radio frequency domain. If stabilized to an atomic transition, an opto-mechanical oscillator can be made to have better long term stability compared to an oven controlled quartz oscillator.



\end{document}